\documentclass[11pt]{article}

\usepackage[a4paper,margin=1in]{geometry}
\usepackage{amsmath,amssymb,bm}
\usepackage{graphicx}
\usepackage{hyperref}
\hypersetup{hidelinks,hypertexnames=false}
\usepackage{booktabs}
\usepackage{microtype}
\usepackage[section]{placeins}
\usepackage{float}

\title{Continuous Event Weighting for Resonance Searches}
\author{
Yi Yang$^{1,2,*}$, Jun-Jie Chang$^{2}$, and Hao-Yin Liao$^{2}$\\
$^{1}$Institute of Physics, Academia Sinica, Taipei 11529, Taiwan\\
$^{2}$Department of Physics, National Cheng Kung University, Tainan 70101, Taiwan\\
$^{*}$\texttt{yiyang429@as.edu.tw}
}
\date{Revised manuscript, November 2026}

\begin{document}
\maketitle

\begin{abstract}
We develop a continuous event weighting framework for resonance searches that uses auxiliary event information without dividing the data into increasingly fine sensitivity categories. A hard selection is a binary weight, finite categories correspond to piecewise constant weights, and in the background dominated limit their fine partition approaches a continuous signal to background density ratio weight. The construction retains auxiliary discrimination in a single physically interpretable resonance spectrum, with calculable statistical variance and direct recovery of the selected signal yield. One million fully explicit event level pseudoexperiments using Gaussian, Gamma, and Beta auxiliary observables verify yield recovery, uncertainty propagation, and approximately nominal frequentist coverage while substantially improving a fitted yield sensitivity proxy. The method is further tested on publicly available CMS Run2016G and Run2016H diphoton Open Data as an independent real data benchmark, using detector level gluon fusion Higgs simulation only as the signal reference. This is a methodological test of the framework, not a CMS Higgs analysis. The results demonstrate a practical route to exploiting continuous auxiliary information without proliferating sensitivity driven event categories.
\end{abstract}

\section{Introduction}

A narrow resonance search is often designed around one privileged observable, denoted here by $m$, such as an invariant mass or recoil mass. The signal is sought as a localized structure in $m$, while many other reconstructed quantities contain additional information that can distinguish signal from background. The experimental problem is therefore not only how to construct a powerful discriminant, but how to use that information without losing the transparency and robustness of the resonance spectrum itself.

A standard solution is to apply selections and then divide the surviving events into mutually exclusive categories with different signal to background ratios, resolutions, or event topologies. This strategy is highly successful. The Higgs discovery at the LHC provides a prominent example: in the diphoton channel, the ATLAS and CMS discovery analyses used mutually exclusive event classes or categories to retain differences in expected sensitivity and mass resolution~\cite{ATLAS:HiggsDiscovery2012,CMS:HiggsDiscovery2012}. Later Higgs measurements developed this strategy further~\cite{ATLAS:Hgg2018,CMS:Hgg2018}. Such categories can serve several purposes. Categories that isolate distinct production modes are essential when separate physics parameters are being measured. Other category boundaries are introduced primarily to retain differences in sensitivity or mass resolution that would be lost in a single inclusive sample.

The latter use raises a simple question. If several categories are eventually combined statistically because they represent different amounts of signal information, is the discretization into categories itself necessary? A hard selection uses only two levels of information, accepted and rejected. A finite category analysis uses several levels. When the categories serve primarily to encode sensitivity, their boundaries are analysis choices rather than physical observables. The natural limiting construction is therefore to let every event retain a continuous amount of information while leaving the resonance coordinate itself unchanged.

Event weighting as a statistical idea is not new. Likelihood based combinations and fractional event counting have long assigned different statistical importance to candidates with different expected signal and background content~\cite{Bock:2007}. Optimal signal to background event weighting for narrow resonance searches was discussed by Freytsis, Ovanesyan, and Thaler~\cite{Freytsis:2009bh}. More recently, Gambhir et al. used learned likelihood ratio weights in a resonance analysis of anti-isolated $\Upsilon$ decays in CMS Open Data~\cite{Gambhir:2025}. Multivariate classifiers also provide continuous discriminants and are standard tools in collider analyses~\cite{Hoecker:2007ht}. A related but distinct construction is the sPlot method~\cite{Pivk:2004ty}, in which fit derived component weights are used to recover the distribution of another variable, subject to the corresponding independence condition. Here the localized structure in the collision data resonance spectrum is not used to determine or optimize the weight. Auxiliary information instead defines a nonnegative sensitivity weight that is applied directly to the resonance spectrum. Variables with smooth explicit mass scaling can be used, but they must be frozen without signal-window information and validated against localized background sculpting. The objective is to improve the sensitivity of the resonance search while preserving a physical yield estimator, rather than to unfold component distributions.

The central contribution is not a new classifier or a new likelihood ratio, but an operational formulation in which auxiliary information is carried directly into a weighted \emph{resonance spectrum}. In this formulation selections and sensitivity driven categories emerge as discrete limiting cases rather than separate analysis concepts. We make explicit four points that are central for a practical search:

\begin{enumerate}
  \item hard selections and finite sensitivity categories are binary and piecewise constant realizations of a common event weight;
  \item the weighted resonance spectrum and its statistical variance have a simple event level form;
  \item a valid weight must not create a localized background structure that can imitate the resonance;
  \item the physical signal yield and its statistical uncertainty remain directly recoverable after weighting.
\end{enumerate}

This viewpoint separates two tasks that are often mixed together. The auxiliary variables decide how much statistical information an event carries, while the resonance variable remains the observable in which the signal is displayed and fitted. The final inference in a precision analysis can still use the profile likelihood and systematic treatment appropriate to that analysis~\cite{Cowan:2010js}. The continuous weight is a way to organize and retain the event information before that final inference.

We first derive the weighted spectrum and its variance. We then show explicitly how selections and categories arise as binary and piecewise constant weights, and how the likelihood ratio emerges in the continuum limit. The method is checked with controlled pseudoexperiments using explicitly specified Gaussian, Gamma, and Beta auxiliary densities and is then demonstrated on real CMS Run2016G and Run2016H Open Data in the $H\to\gamma\gamma$ channel~\cite{CMS:DoubleEGRunGOpenData,CMS:DoubleEGRunHOpenData,LassilaPerini:2016,Rizzi:2019}. The known Higgs resonance provides a stringent real detector test of the full workflow without turning the study into a Higgs rate measurement. The broader target is a future resonance search in which one or a small number of detector level signal simulations and data driven background information can be converted into an information efficient but still transparent resonance analysis.

\section{Continuous event weighting}
\label{sec:formalism}

\subsection{Weighted resonance spectrum}
\label{sec:spectrum}

Let $P(m,x)$ denote the expected joint event density in the resonance variable $m$ and the auxiliary variables $x$. We write
\begin{equation}
  P(m,x)=N_S f_S(m)p_S(x|m)+N_B f_B(m)p_B(x|m),
  \label{eq:joint}
\end{equation}
where $N_S$ and $N_B$ are the physical signal and background yields. The functions $f_S(m)$ and $f_B(m)$ are normalized probability densities of the resonance variable, while $p_S(x|m)$ and $p_B(x|m)$ describe the auxiliary variables $x$ conditional on $m$.

Each selected event is assigned a nonnegative weight $w(x)$. No independence between $x$ and $m$ is assumed in Eq.~\eqref{eq:joint}. For a bump hunt, however, the weight must not be tuned using the localized signal structure in $m$; any auxiliary observable with explicit or implicit mass dependence must be frozen independently and pass the no sculpting test below. Define
\begin{equation}
  \bar w_i(m)=\int dx\,w(x)p_i(x|m),
  \qquad
  w_{2,i}(m)=\int dx\,w^2(x)p_i(x|m),
  \qquad i=S,B.
  \label{eq:weightmoments}
\end{equation}
The individual event weights are $w_k=w(x_k)$, whereas the bars in Eq.~\eqref{eq:weightmoments} denote component averages. Define the random weighted event measure $N_w(m)=\sum_k w(x_k)\delta(m-m_k)$; its integral over a mass bin is the observed sum of weights. For a binned spectrum, the observed weighted content and squared weight sum in bin $j$ are
\begin{equation}
  y_j^w=\sum_{k\in j}w_k,
  \qquad
  q_j^w=\sum_{k\in j}w_k^2.
  \label{eq:binsums}
\end{equation}
The expectation value of the continuous weighted spectrum is
\begin{equation}
  \mathbb{E}[N_w(m)]
  =N_S f_S(m)\bar w_S(m)
  +N_B f_B(m)\bar w_B(m).
  \label{eq:master}
\end{equation}
Equation~\eqref{eq:master} is the basic relation of the method. It does not require the auxiliary variables to be independent and it does not require the weight to be an exact likelihood ratio.

\subsection{Background shape and statistical variance}
\label{sec:moments}

The weighted background is proportional to
\begin{equation}
  f_B^{(w)}(m)\propto f_B(m)\bar w_B(m).
  \label{eq:weightedbackground}
\end{equation}
The requirement is therefore not that the weighted background be identical to the unweighted background. Smooth changes of slope or normalization are allowed and can be fitted. The essential condition is that weighting must not generate a localized structure that is degenerate with the signal being searched for. We refer to this as the no sculpting requirement. In a data analysis it should be tested directly in sidebands or control samples. The practical validation and uncertainty treatment are discussed in Sec.~\ref{sec:systematics}.

For a Poisson point process with a fixed weight function, the variance of a weighted sum is the expectation of the squared weight sum; the observed $q_j^w$ is an unbiased estimator of this variance~\cite{BohmZech:2014}. In a local mass window $\mathcal W$, let $S_{\mathcal W}$ and $B_{\mathcal W}$ be the expected unweighted yields. In the following local formulas, $\bar w_i$ and $w_{2,i}$ denote the corresponding component averages over events in that window. The first and second moment relations are then exact:
\begin{align}
  S_{w,\mathcal W}&=S_{\mathcal W}\bar w_S,
  &\sigma^2_{S,\mathcal W}&=S_{\mathcal W} w_{2,S},\\
  B_{w,\mathcal W}&=B_{\mathcal W}\bar w_B,
  &\sigma^2_{B,\mathcal W}&=B_{\mathcal W} w_{2,B}.
  \label{eq:meanvariance}
\end{align}
These identities do not require Gaussian distributed observables or Gaussian event sums.

Search sensitivity can be quantified by an expected or median local discovery significance~\cite{Cowan:2010js}. For the analytic optimization below, we use the simpler linear figure of merit, defined by
\begin{equation}
  Z_w^{\rm lin}\equiv\frac{S_{\mathcal W}\bar w_S}{\sqrt{B_{\mathcal W}w_{2,B}}}.
  \label{eq:Zw}
\end{equation}
This approximates discovery significance in the background dominated, locally Gaussian regime. The moment identities themselves do not require Gaussianity. Relative to $Z_0^{\rm lin}=S_{\mathcal W}/\sqrt{B_{\mathcal W}}$, the gain in this defined figure of merit is exactly
\begin{equation}
  R_Z[w]\equiv\frac{Z_w^{\rm lin}}{Z_0^{\rm lin}}
  =\frac{\bar w_S}{\sqrt{w_{2,B}}}.
  \label{eq:RZ}
\end{equation}
A spectrum can look much cleaner after aggressive weighting while having worse statistical sensitivity if the squared weights are dominated by a few events.

\subsection{Selected signal yield and uncertainty propagation}
\label{sec:yield}

Weighting changes the displayed signal normalization but need not change the parameter being measured. Over a fitted mass range $R$, let $\kappa_S$ denote the mean signal weight integrated over the normalized signal mass shape,
\begin{equation}
  \kappa_S=\int_R dm\, f_S(m)\bar w_S(m).
  \label{eq:kappas}
\end{equation}
If a fit uses a signal shape normalized to unit weighted yield, the fitted weighted excess is $Y_S^w=N_S\kappa_S$ and
\begin{equation}
  \widehat N_S=\frac{\widehat Y_S^w}{\kappa_S},
  \qquad
  {\rm Var}(\widehat N_S)
  =\frac{{\rm Var}(\widehat Y_S^w)}{\kappa_S^2},
  \label{eq:yieldrecovery}
\end{equation}
for fixed $\kappa_S>0$. A finite simulation or control sample uncertainty on $\kappa_S$ is an additional calibration uncertainty and can be introduced as a nuisance parameter.

An equivalent implementation, used in the numerical studies below, avoids a separate conversion. For mass bin $j$, let $t_{S,j}^{(1)}$ and $t_{S,j}^{(2)}$ denote, respectively, the first and second weight moment signal templates per physical selected signal event,
\begin{equation}
  t_{S,j}^{(1)}=\int_j dm\,f_S(m)\bar w_S(m),
  \qquad
  t_{S,j}^{(2)}=\int_j dm\,f_S(m)w_{2,S}(m).
  \label{eq:weightedtemplates}
\end{equation}
For bins covering $R$, $\sum_j t_{S,j}^{(1)}=\kappa_S$: a bin template is a contribution to the integrated mean weight, rather than that entire mean. The background templates $t_{B,j}^{(1,2)}$ are defined by the same integrals with $S$ replaced by $B$. The expected weighted signal in that bin is then $N_S t_{S,j}^{(1)}$, and its contribution to the weighted variance is $N_S t_{S,j}^{(2)}$. The fitted coefficient is directly $N_S$, expressed in selected event units even when $\bar w_S(m)$ varies across the fit range.

In the high count binned fit used below, the statistical model needs both the predicted weighted content and its variance. Although $q_j^w$ is an unbiased variance estimator, inserting its fluctuating value into an inverse variance can correlate the fit weight with the same data fluctuation being fitted. We instead model the expected variance. In a two component Poisson model,
\begin{equation}
  V_j(N_S,N_B)
  =N_S t_{S,j}^{(2)}+N_B t_{B,j}^{(2)}.
  \label{eq:modelvariance}
\end{equation}
Equation~\eqref{eq:modelvariance} is an exact second moment relation for the fixed-template Poisson model; it is not conditional on a Gaussian approximation. Approximating the distribution of the weighted bin contents by Gaussians is a separate step. The controlled pseudoexperiments below test the fitted selected signal yield and its propagated statistical uncertainty directly.

\subsection{Likelihood construction and systematic uncertainties}
\label{sec:systematics}

Weighted Poisson event sums follow a compound Poisson distribution~\cite{BohmZech:2014}. Their first and second moment identities are exact. For the high count fit used here, we approximate independent weighted mass bins by Gaussians. With nuisance parameters $\theta$ and a background family index $d$, write
\begin{align}
 \nu_j&=N_S t_{S,j}^{(1)}(\theta)+N_B t_{B,j}^{(1)}(\theta,d),\\
 V_j&=N_S t_{S,j}^{(2)}(\theta)+N_B t_{B,j}^{(2)}(\theta,d),
 \label{eq:nuisancemoments}\\
 -2\ln L_w&=\sum_j\left[\frac{(y_j^w-\nu_j)^2}{V_j}+\ln V_j\right]
 -2\ln C(\theta)+\text{constant},
 \label{eq:weightedlikelihood}
\end{align}
where $C$ contains auxiliary calibration constraints. The parameter-dependent $\ln V_j$ term is retained. At low counts, the compound Poisson law, another suitable approximation, or event level pseudoexperiments can be used instead.

Standard detector and signal-model uncertainties can be propagated by applying the frozen weight function to reconstructed simulated events with varied scales, resolutions, identification, or signal modeling. Each variation changes both moment templates; common sources use common nuisance parameters. The calibration $\kappa_S=\sum_jt_{S,j}^{(1)}$ follows from the same templates, while the observed weighted data remain fixed. Finite simulation and density-estimation uncertainties can be assessed by resampling the construction samples. If training data are reused in the fit, the calibration must also account for the learning procedure. The ensembles below test the statistical errors conditional on the frozen models.

Background-family uncertainty can be treated by discrete profiling~\cite{Dauncey:2015}: profile $d$ together with the background parameters and an appropriate complexity penalty, validating the interval calibration for the weighted model. A simultaneous fit to sidebands and the search region is also possible, with each observation included once.

To assess sculpting, apply the frozen weighting to suitable background control samples or plausible alternative background models and fit the resulting background-only spectra with the proposed signal plus background model, as in spurious-signal studies~\cite{ATLAS:Hgg2014}. Let $N_S^{\rm spur}$ denote the mean fitted signal, or the fit to a high statistics expected spectrum, accounting for validation sample uncertainty and allowing a signed signal estimate. An analysis can require
\begin{equation}
 \frac{|N_S^{\rm spur}|}{\sigma_{N_S}^{\rm stat}}<\epsilon_{\rm bias},
 \label{eq:spuriouscriterion}
\end{equation}
over the tested mass hypotheses and backgrounds, with an analysis-specific tolerance. At fixed tolerance, the permitted fractional background discrepancy decreases as statistical precision improves, making this check particularly relevant at high luminosity. Smooth sidebands or small mass-score correlations alone do not bound bias inside a masked window.

One way to cover a residual bias is
\begin{equation}
 \nu_j\longrightarrow\nu_j+\eta_{\rm sp}\Delta_{\rm sp}t_{S,j}^{(1)},\qquad
 C\longrightarrow C\exp(-\eta_{\rm sp}^2/2),
 \label{eq:spuriousnuisance}
\end{equation}
where $\Delta_{\rm sp}$ is a signal-equivalent yield envelope in selected event units, including validation uncertainty. This models uncertain background projection, not additional physical signal events in $V_j$; physical background variations also change the second moment. The same uncertainty should not be added again if it is already covered by the profiled background model. These are alternative or complementary treatments, rather than requirements to use all of them.

\section{Selections and categories as discrete weights}
\label{sec:categories}

The central observation of this work is that selection, categorization, and continuous weighting can be written in one language.

A hard selection on a region $\mathcal A$ of the auxiliary space $x$ is the binary weight
\begin{equation}
  w_{\rm cut}(x)=
  \begin{cases}
    1, & x\in \mathcal A,\\
    0, & x\notin \mathcal A.
  \end{cases}
  \label{eq:cutweight}
\end{equation}
All information inside the accepted region is treated equally, while all information outside it is discarded.

Now partition the auxiliary space into mutually exclusive categories $C_k$. A category based linear combination is equivalent to the piecewise constant weight
\begin{equation}
  w_{\rm cat}(x)=\sum_k a_k\,\mathbf{1}_{C_k}(x),
  \label{eq:catweight}
\end{equation}
Let $S_k$ and $B_k>0$ be the expected signal and background in category $k$ within the same mass window $\mathcal W$. The indicator $\mathbf{1}_{C_k}$ is one inside $C_k$ and zero elsewhere. The linear figure of merit is
\begin{equation}
  Z_{\rm cat}^{\rm lin}
  \equiv
  \frac{\sum_k a_kS_k}
       {\sqrt{\sum_k a_k^2B_k}}.
  \label{eq:Zcat}
\end{equation}
To apply Cauchy Schwarz, set $u_k=a_k\sqrt{B_k}$ and $v_k=S_k/\sqrt{B_k}$. Then
\[
 (\sum_k a_kS_k)^2=(\sum_k u_kv_k)^2
 \le (\sum_k a_k^2B_k)(\sum_k S_k^2/B_k).
\]
Equality requires $u_k\propto v_k$, giving
\begin{equation}
  a_k\propto\frac{S_k}{B_k},
  \qquad
  (Z_{\rm cat,max}^{\rm lin})^2=\sum_k\frac{S_k^2}{B_k}.
  \label{eq:catopt}
\end{equation}
A finite category based linear combination is therefore a discretized weighting of the auxiliary space. As the categories are refined, their yield ratios approach a constant times the local auxiliary density ratio.

To connect this local argument with Eq.~\eqref{eq:joint}, define the normalized auxiliary density in the same mass window $\mathcal W$ by
\begin{equation}
  p_i^{\mathcal W}(x)=
  \frac{\displaystyle\int_{\mathcal W} dm\,f_i(m)p_i(x|m)}
       {\displaystyle\int_{\mathcal W} dm\,f_i(m)},
  \qquad i=S,B.
  \label{eq:windowdensity}
\end{equation}
Writing $F_i=\int_{\mathcal W}f_i(m)\,dm$, direct integration of Eq.~\eqref{eq:joint} gives $S_{w,\mathcal W}=N_S F_S\int w(x)p_S^{\mathcal W}(x)\,dx$ and $\sigma^2_{B,\mathcal W}=N_B F_B\int w^2(x)p_B^{\mathcal W}(x)\,dx$. Thus the window averages in Eq.~\eqref{eq:RZ} use precisely these normalized densities, without an independence assumption. Below we write $p_i(x)$ for $p_i^{\mathcal W}(x)$ to keep the notation compact. For $p_B>0$ on the signal support and a finite integral on the right, applying Cauchy Schwarz to $w\sqrt{p_B}$ and $p_S/\sqrt{p_B}$ gives
\begin{equation}
  R_Z[w]
  \le
  \left[\int dx\,\frac{p_S^2(x)}{p_B(x)}\right]^{1/2},
  \label{eq:continuousbound}
\end{equation}
with equality for
\begin{equation}
  w(x)\propto r(x)
  \equiv\frac{p_S(x)}{p_B(x)}.
  \label{eq:lr}
\end{equation}
This is the same likelihood ratio that appears in the Neyman Pearson construction~\cite{NeymanPearson:1933}. Its role as an optimal linear weight for narrow resonance searches was also derived in Ref.~\cite{Freytsis:2009bh}. Here it connects the category optimum in Eq.~\eqref{eq:catopt} with the continuous bound in Eq.~\eqref{eq:continuousbound}, for the same mass window and background dominated linear criterion. The density ratio need not be estimated by explicit multidimensional histograms. A calibrated probabilistic classifier, including a boosted decision tree or neural network, can provide the same information. If $D(x)=P_{\rm train}(S|x)$ is the calibrated binary classifier posterior and $\pi_S=P_{\rm train}(S)$, $\pi_B=P_{\rm train}(B)$ are its training class fractions, with $\pi_S+\pi_B=1$, Bayes' theorem gives
\begin{equation}
  \frac{p_S(x)}{p_B(x)}
  =\frac{\pi_B}{\pi_S}\frac{D(x)}{1-D(x)}.
  \label{eq:classifierlr}
\end{equation}
This classifier to likelihood ratio connection is well known~\cite{Cranmer:2015bka}. The categories $C_k$ can be intervals of this classifier score or any partition of $x$; they are not the training classes or their priors. A monotonic classifier score is already sufficient to define selections or categories, while using it as a quantitative event weight requires an appropriate calibration or density ratio estimate.

Equations~\eqref{eq:catopt} and \eqref{eq:continuousbound} make the connection precise. A hard cut is a two level use of auxiliary information. Finite categorization is a piecewise constant approximation. The continuous likelihood ratio is the limiting form when the partition becomes arbitrarily fine. This statement concerns the sensitivity weighting role of categories for a common signal hypothesis. It is not an algebraic identity with an arbitrary multi category profile likelihood whose categories have different signal shapes, resolutions, or nuisance parameter structures. Categories introduced to measure distinct production modes or separate physics parameters have an additional role and are not replaced by a single inclusive weight.

The exact ratio in Eq.~\eqref{eq:lr} is a population quantity. In a low dimensional application it can be estimated with joint histograms or density models from signal simulation and background control data. In higher dimensions, a calibrated classifier provides a practical estimate without explicit fine multidimensional bins. Finite sample fluctuations, calibration, and poorly populated regions affect either approach; they motivate validation and regularization before the signal region is examined. The estimated likelihood ratio is not always the most stable practical weight. We therefore use the simple family
\begin{equation}
  w_\alpha(x)=r^\alpha(x)
  \label{eq:poweredweight}
\end{equation}
when tempering is useful. Here $\alpha=0$ gives the unweighted sample and $\alpha=1$ gives the likelihood ratio weight. Values $0<\alpha<1$ compress the most extreme likelihood ratios, reducing the influence of sparsely populated tails and density modeling fluctuations. In the ideal background dominated limit the exact ratio is optimal, but in a finite analysis a tempered choice can retain nearly the same separation with larger effective statistics and more stable control sample behavior. There is no universal preferred tempered exponent. The value of $\alpha$ should be fixed using independent simulation or control data before examining the signal region.

The effective weighted sample size
\begin{equation}
  N_{\rm eff}=\frac{(\sum_k w_k)^2}{\sum_k w_k^2}
  \label{eq:neff}
\end{equation}
provides a direct diagnostic of long weight tails. Correlated auxiliary variables should be described by a joint density ratio or a multivariate discriminant. Multiplying one dimensional ratios is justified only when the relevant factorization is an adequate approximation.

\section{Controlled event level validation}
\label{sec:controlled}

We first test the statistical relations in a fully specified model that is independent of detector data. Each pseudoexperiment contains Poisson fluctuated mean yields of $10^6$ background events and $10^3$ signal events. The resonance coordinate is generated independently of the auxiliary observables. The signal mass distribution is a Gaussian with mean 91.2~GeV and width 1.5~GeV, truncated to $70<m<110$~GeV. The background follows
\begin{equation}
  f_B(m)\propto e^{-0.025(m-70\,{\rm GeV})}
  \qquad (70<m<110~{\rm GeV}).
  \label{eq:toymass}
\end{equation}

\begin{figure}[H]
  \centering
  \includegraphics[width=0.96\textwidth]{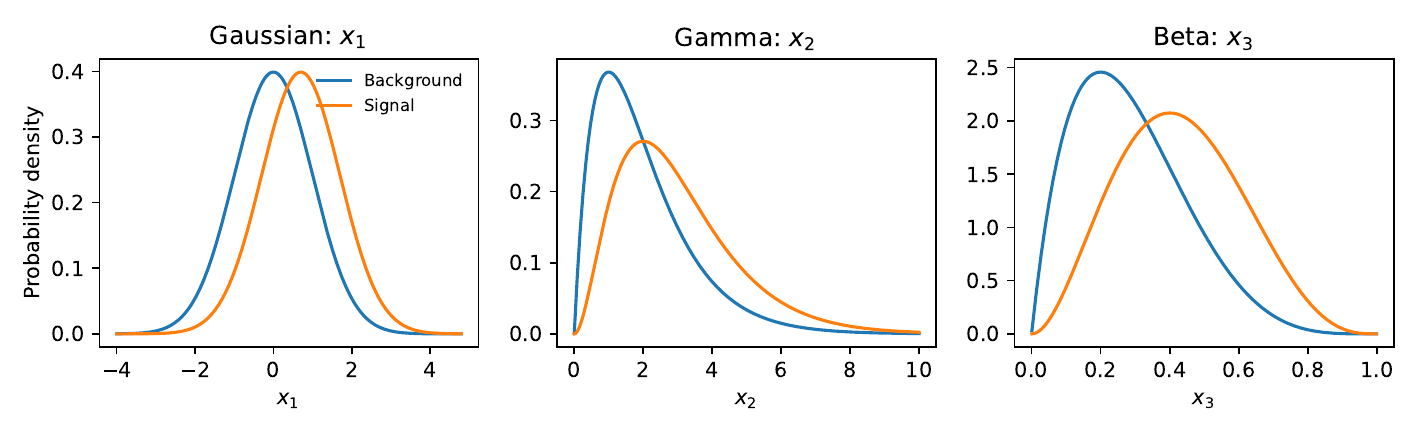}
  \caption{Auxiliary distributions used in the controlled pseudoexperiments. The three inputs have Gaussian, positively skewed Gamma, and bounded Beta forms. Each variable provides moderate signal to background discrimination, and the event weight uses their exact joint likelihood ratio.}
  \label{fig:auxexamples}
\end{figure}

Every event has three independent auxiliary observables with deliberately different functional forms,
\begin{align}
 x_1^B&\sim {\cal N}(0,1),&
 x_1^S&\sim {\cal N}(0.7,1),
 \label{eq:toygaussian}\\
 x_2^B&\sim {\rm Gamma}(2,1),&
 x_2^S&\sim {\rm Gamma}(3,1),
 \label{eq:toygamma}\\
 x_3^B&\sim {\rm Beta}(2,5),&
 x_3^S&\sim {\rm Beta}(3,4).
 \label{eq:toybeta}
\end{align}
Thus the demonstration is not restricted to Gaussian inputs. Figure~\ref{fig:auxexamples} shows the actual one dimensional densities used in the pseudoexperiments. The exact joint likelihood ratio is
\begin{equation}
 r(x_1,x_2,x_3)=r_1(x_1)r_2(x_2)r_3(x_3),
 \label{eq:toyweight}
\end{equation}
with
\begin{equation}
 r_1=e^{0.7x_1-0.245},\qquad
 r_2=\frac{x_2}{2},\qquad
 r_3=\frac{2x_3}{1-x_3}.
 \label{eq:toyfactors}
\end{equation}
The baseline controlled test uses $w=r$. Here $E_i$ denotes an average over the auxiliary density $p_i$, so $E_B[w^2]=w_{2,B}$ and $E_S[w]=\bar w_S$. Specifically for the exact ratio $r=p_S/p_B$, $E_S[r^q]=E_B[r^{q+1}]$. The exact moments are $E_B[w]=1$, $E_B[w^2]=E_S[w]=4.897$, and $E_B[w^3]=E_S[w^2]=104.38$. Equation~\eqref{eq:RZ} therefore predicts a background dominated sensitivity gain of $\sqrt{4.897}=2.21$ before details of the mass fit are included.

\begin{figure}[H]
  \centering
  \includegraphics[width=0.94\textwidth]{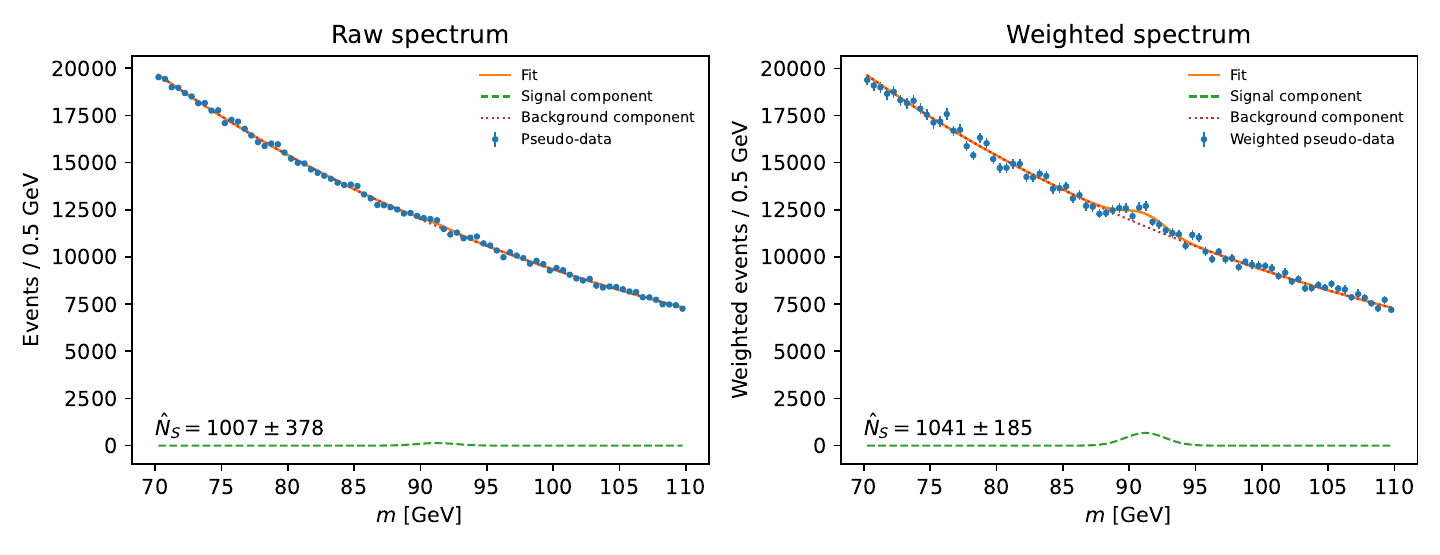}
  \caption{Representative fully explicit event level pseudoexperiment in the controlled mixed distribution model. All three auxiliary variables are generated for every event, and the weight does not use the resonance variable.}
  \label{fig:controlled}
\end{figure}

A representative experiment is shown in Fig.~\ref{fig:controlled}. Both spectra are fitted in 0.5~GeV bins with a linear generalized least squares estimator for the selected signal yield and background normalization. With $X=(t_S^{(1)},t_B^{(1)})$ and a fixed background-only reference covariance $V_0$, the estimator and its covariance are
\begin{equation}
 A=(X^{T}V_0^{-1}X)^{-1}X^{T}V_0^{-1},\qquad
 \widehat{\boldsymbol N}=A\boldsymbol y,\qquad
 {\rm Cov}(\widehat{\boldsymbol N})=A\,\operatorname{diag}(V_j)A^{T}.
 \label{eq:controlledfit}
\end{equation}
The covariance is evaluated using Eq.~\eqref{eq:modelvariance} with the fitted components; a negative fitted component is set to zero for this covariance calculation. No detector or background model systematic uncertainties are included in this controlled test. The quoted yield uncertainty is the square root of the signal diagonal covariance entry. Since mass and auxiliary observables are independent here, the signal mean weight is one global calibration, $E_S[w]$, multiplying the normalized signal mass template in every bin. The second moment enters the variance template. The reported $\widehat N_S/\sigma(\widehat N_S)$ is a fitted yield sensitivity proxy, distinct from the likelihood ratio diagnostic used in the CMS example.

The validation is then repeated for one million statistically independent pseudoexperiments. Every background and signal event, including its mass and three auxiliary observables, is generated explicitly; background generation is processed in memory limited chunks. Table~\ref{tab:controlled} summarizes the ensemble. The raw and weighted mean fitted yields are 999.80 and 999.77 for an injected mean of 1000. The ensemble RMS values, 378.14 and 184.70 events, agree with the corresponding mean fitted uncertainties, 378.37 and 184.97 events. The median fitted yield sensitivity increases from 2.642 to 5.398. The improvement in precision and in the fitted yield sensitivity proxy are both about a factor of two. The observed RMS ratio, 2.05, differs from the background-only gain 2.21 because the actual uncertainty includes the signal variance term $N_S t_{S,j}^{(2)}$, enhanced by $E_S[w^2]=104.38$. With the same background normalization fit, the model covariance predicts standard deviations 378.37 and 184.99 events, a gain 2.045 consistent with the observed RMS ratio.

\begin{table}[H]
  \centering
  \caption{Controlled validation from $10^6$ fully explicit pseudoexperiments. Coverage is defined with respect to the fixed injected mean signal yield, not the Poisson realized signal count in each pseudoexperiment.}
  \label{tab:controlled}
  \begin{tabular}{lcc}
    \toprule
    Quantity & Raw & Continuous weight \\
    \midrule
    Mean fitted $N_S$ & 999.80 & 999.77 \\
    Ensemble RMS & 378.14 & 184.70 \\
    Mean fitted $\sigma(N_S)$ & 378.37 & 184.97 \\
    Median $N_S/\sigma(N_S)$ & 2.642 & 5.398 \\
    Pull mean & $-0.0019$ & $-0.0150$ \\
    Pull width & 0.9994 & 0.9989 \\
    $1\sigma$ coverage & 0.6832 & 0.6843 \\
    95\% coverage & 0.9502 & 0.9501 \\
    \bottomrule
  \end{tabular}
\end{table}

\begin{figure}[H]
  \centering
  \includegraphics[width=0.94\textwidth]{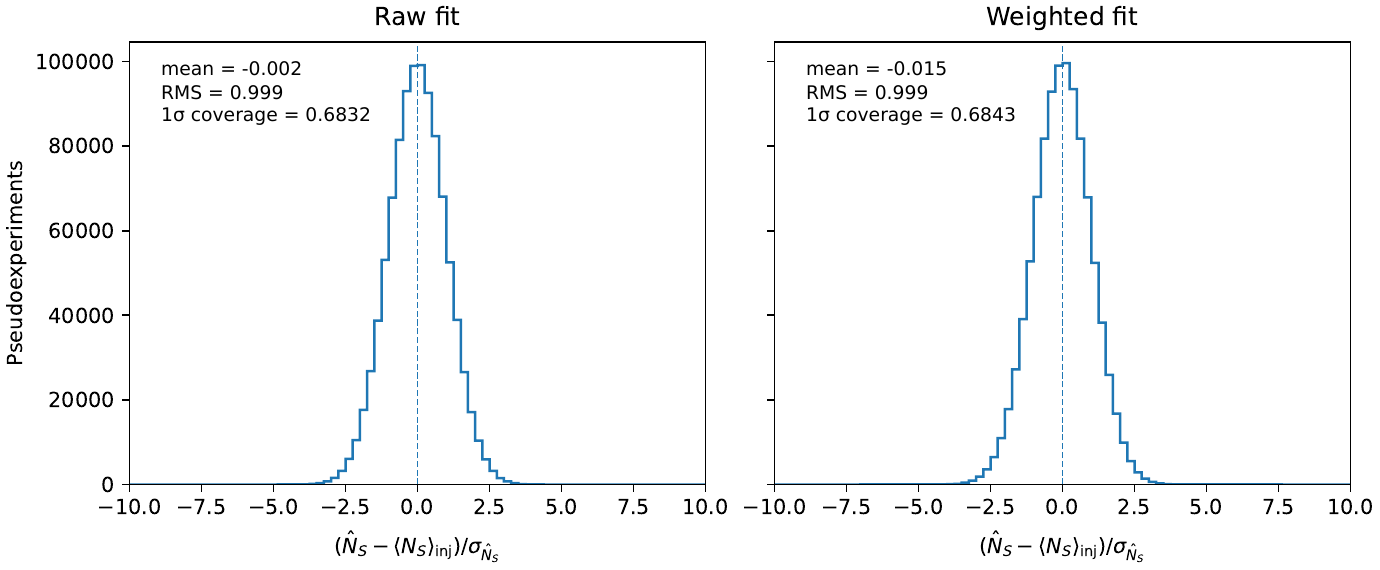}
  \caption{Pull distributions from one million fully explicit controlled pseudoexperiments. Both panels use the common range $-10$ to $+10$. The fitted uncertainties reproduce the ensemble fluctuations and approximately nominal coverage for both the raw and weighted spectra.}
  \label{fig:controlpulls}
\end{figure}

Figure~\ref{fig:controlpulls} shows the pull distributions,
\begin{equation}
  {\rm pull}=\frac{\widehat N_S-\langle N_S\rangle_{\rm inj}}
  {\sigma_{\widehat N_S}}.
  \label{eq:pull}
\end{equation}
Coverage is the fraction of trials satisfying $|\widehat N_S-N_{S,\rm inj}|\le\sigma_{\widehat N_S}$ or $1.96\sigma_{\widehat N_S}$, with Gaussian nominal probabilities 68.27\% and 95.00\%. The weighted $1\sigma$ coverage 0.6843 shows a small overcoverage. The near-unit pull widths and the agreement of ensemble RMS with fitted uncertainties support the uncertainty approximation at the precision shown.

The auxiliary variables are independent in this controlled baseline so that the distinct distributional forms can be inspected transparently. Correlation is not assumed by the formalism; when present, it belongs in the joint density ratio or an appropriately calibrated multivariate discriminant.

\section{CMS Open Data demonstration}
\label{sec:cms}

We test the method on publicly available CMS Run2016G and Run2016H DoubleEG Open Data, which are real proton proton collision data, with public 2016 gluon fusion $H\to\gamma\gamma$ detector level MC used only to construct the signal auxiliary density and signal templates~\cite{CMS:DoubleEGRunGOpenData,CMS:DoubleEGRunHOpenData,CMS:HggOpenMC,Rizzi:2019}. This is an independent methodological test using CMS Open Data, not a CMS Higgs analysis or a reproduction of the official CMS event categorization. The two data periods are always used together. No official $H\to\gamma\gamma$ analysis categories are imported: the starting point is one baseline selected diphoton sample, and the analysis defined score bins introduced later are used only for the controlled method comparison. The signal reference is detector level MC rather than truth level MC. The strategy is to learn signal auxiliary distributions and mass templates from this detector level signal MC, and to learn the background auxiliary distributions and smooth mass model from collision data sidebands. The continuum background is therefore data driven rather than taken from background MC. Both NanoAOD samples provide reconstructed photon kinematics and identification quantities. The quoted event counts are after the selection below. The known Higgs mass sets a masked collision data window for this standard candle test; events in that window are not used to construct or choose the weight. Signal simulation supplies the signal reference independently of any excess in the collision data.

The collision data are restricted to certified luminosity sections and the same available diphoton trigger OR is applied to data and signal simulation. Photon candidates satisfy $p_T>20$~GeV, $|\eta|<2.5$ excluding the barrel endcap transition, the electron veto, and the loose NanoAOD cut based photon selection. The two leading accepted photons must satisfy
\begin{equation}
  p_T^{\gamma_1}/m_{\gamma\gamma}>1/3,
  \qquad
  p_T^{\gamma_2}/m_{\gamma\gamma}>1/4,
  \label{eq:cmsselection}
\end{equation}
with $100<m_{\gamma\gamma}<180$~GeV. After these requirements the combined sample contains 337801 events, of which 236318 lie in the sidebands $100$--$115$ and $135$--$180$~GeV. The selected data counts are 155328 for Run2016G and 182473 for Run2016H. The signal reference contains 300000 selected detector level gluon fusion MC events and is used consistently to define the empirical signal density and signal templates. No Higgs cross section or rate inference is attempted from this MC reference. The collision data mass window $115$--$135$~GeV is excluded from the weight construction and pre-unblinding validation. This masking does not exclude the corresponding signal MC events, which define the independently supplied signal hypothesis.

\subsection{Auxiliary variables, correlation, and blinded weight choice}
\label{sec:cmsweight}

\begin{figure}[H]
  \centering
  \includegraphics[width=0.98\textwidth]{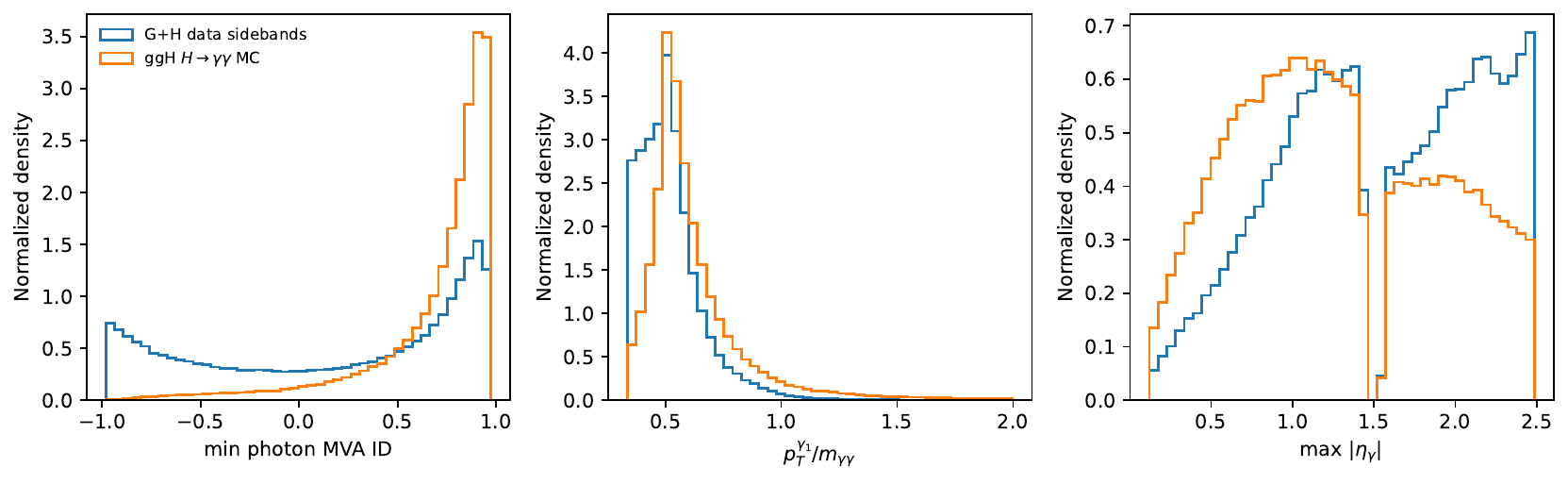}
  \caption{Normalized CMS benchmark inputs. Background is represented by the combined Run2016G and Run2016H data sidebands, while signal is the selected public gluon fusion $H\to\gamma\gamma$ simulation. The maximum photon pseudorapidity is shown to document detector topology; it is not multiplied into the final likelihood ratio as an independent continuous factor.}
  \label{fig:cmsauxvars}
\end{figure}

Three transparent collider observables were examined,
\begin{equation}
  u=\min({\rm Photon\ MVAID}),\qquad
  v=p_T^{\gamma_1}/m_{\gamma\gamma},\qquad
  e=\max|\eta_\gamma|.
  \label{eq:cmsvars}
\end{equation}
The minimum and maximum run over the two selected photons, and $\gamma_1$ denotes the leading photon in $p_T$. Here ${\rm Photon\ MVAID}$ is a reconstructed photon identification score stored in NanoAOD; it is not a CMS Higgs event category label.
Their sideband and signal simulation distributions are shown in Fig.~\ref{fig:cmsauxvars}. They are visibly non Gaussian and provide different amounts of discrimination. The $e$ distribution also displays the detector topology directly. The ratio $v=p_T^{\gamma_1}/m_{\gamma\gamma}$ contains the resonance mass as a smooth kinematic scale and is therefore not algebraically independent of $m_{\gamma\gamma}$. It is not optimized on the Higgs peak: its density model and the exponent are fixed using sidebands and signal simulation, and the resulting weighted to unweighted mass dependence is explicitly validated for smoothness before unblinding.

Before unblinding, the data sidebands and signal MC are each split 50:50 into a construction subset and a held out validation subset. A first factorized construction used
\[
 r_{\rm 3fac}(u,v,e)=
 \frac{p_S(u)p_S(v)p_S(e)}{p_B(u)p_B(v)p_B(e)}.
\]
Photon identification response can depend on detector region. To retain this dependence explicitly, a detector region conditioned alternative was compared with $r_{\rm 3fac}$ using the same construction and held out subsets. This is the sense in which the alternative is more conservative: it does not treat the continuous detector topology observable as independent of the identification response. Define $c=0$ for events with both photons in the barrel, $e<1.4442$, and $c=1$ for selected events containing at least one endcap photon. The final auxiliary ratio is
\begin{equation}
  r(c,u,v)=
  \frac{P_S(c)}{P_B(c)}
  \frac{p_S(u|c)}{p_B(u|c)}
  \frac{p_S(v|c)}{p_B(v|c)}.
  \label{eq:cmsratio}
\end{equation}
For each $c$ separately, $u$ and $v$ are histogrammed using 30 intervals whose boundaries are empirical quantiles of the construction background sample in that region. Thus these are approximately equal-background-population bins in the continuous variable, not bins in $c$. The signal histogram uses the same edges. Adding 0.5 to each bin count before normalizing the histogram supplies the pseudocount and prevents empty-bin ratios; the bin widths cancel between signal and background in each ratio. Region probabilities $P_i(c)$ are estimated from their event fractions with the same 0.5-count smoothing. This factorization is a deliberately transparent approximation, not a requirement of the method. The largest absolute pairwise Pearson coefficients in the held out three variable sample are 0.187 for background and 0.135 for signal. Small pairwise coefficients do not prove exact factorization; a joint density or calibrated classifier can be substituted in a higher dimensional application. Neglecting a correlation can reduce separation without intrinsically biasing the yield: the formalism applies to any fixed weight if its actual signal moment templates and weighted background are modeled correctly. Bias can arise if the resulting mass dependence or signal calibration is also mismodeled.

The continuous weighting formalism itself does not require a fitted category boundary or a tunable exponent: $\alpha=1$ is the formal likelihood ratio choice. In this finite sample benchmark, however, we also examine two optional regularization choices, detector region conditioning and tempering with $\alpha<1$. The construction and validation subsets just defined are used to choose these implementation details without the Higgs window. Table~\ref{tab:alphascan} shows the region conditioned scan on the held out subset. The choice $\alpha=0.75$ maximizes the held out sensitivity proxy in the scanned set. The untempered value $\alpha=1$ gives nearly the same proxy but a smaller background effective sample size and larger closure residuals. The value $\alpha=0.75$ is therefore frozen as a finite sample stability compromise, not as a universal optimum. At the same exponent, $\alpha=0.75$, the original $r_{\rm 3fac}(u,v,e)$ construction gives $\widehat R_Z=1.503$, $N_{\rm eff}/N=0.411$, and a closure RMS of 2.53\%, all less favorable than the region conditioned result. This 50:50 split is not intrinsic to the method and does not discard data from the final analysis: after the architecture and exponent are frozen, all sideband events are reused to build the final density ratio. In a search where these implementation choices are fixed from background simulation or independent control samples, no collision data split is required.

For clarity, the two table diagnostics are defined explicitly. The proxy is the empirical version of Eq.~\eqref{eq:RZ},
\begin{equation}
 \widehat R_Z=\frac{\langle w\rangle_{S,\rm val}}
 {\sqrt{\langle w^2\rangle_{B,\rm val}}}.
 \label{eq:cmsproxy}
\end{equation}
Here background averages use held out sideband events, so this proxy does not include a mass fit or establish the background distribution inside the masked window. For 2~GeV mass bins, let $n_j$ and $y_j^w$ be the held out sideband unweighted and weighted contents, and fit their ratio $\rho_j=y_j^w/n_j$ with a quadratic $g_2(m_j)$ using sideband bins only. Then
\begin{equation}
 \delta_j=\frac{\rho_j}{g_2(m_j)}-1,\qquad
 {\rm Closure\ RMS}=\sqrt{\frac{1}{J_{\rm SB}}\sum_{j\in{\rm SB}}\delta_j^2},
 \label{eq:closure}
\end{equation}
and the maximum residual is $\max_{j\in{\rm SB}}|\delta_j|$. The quadratic is a descriptive smoothing fit, with fitting weights proportional to $n_j/\sqrt{q_j^w}$. The reported RMS includes statistical fluctuations; it is neither a goodness of fit probability nor a bound on signal yield bias. All mentions of the held out sample in this subsection refer to this same validation split.

\begin{table}[H]
  \centering
  \caption{Held out pre-unblinding stability study for the region conditioned CMS weight. The closure quantities are descriptive residual measures after a smooth sideband fit, not goodness of fit probabilities.}
  \label{tab:alphascan}
  \begin{tabular}{ccccc}
    \toprule
    $\alpha$ & $\widehat R_Z$ proxy & $N_{\rm eff}^B/N_B$ & Closure RMS & Max. residual \\
    \midrule
    0.50 & 1.459 & 0.667 & 1.56\% & 4.23\% \\
    0.75 & 1.519 & 0.502 & 2.06\% & 5.60\% \\
    1.00 & 1.517 & 0.372 & 2.54\% & 7.31\% \\
    \bottomrule
  \end{tabular}
\end{table}

\begin{figure}[t]
  \centering
  \includegraphics[width=0.72\textwidth]{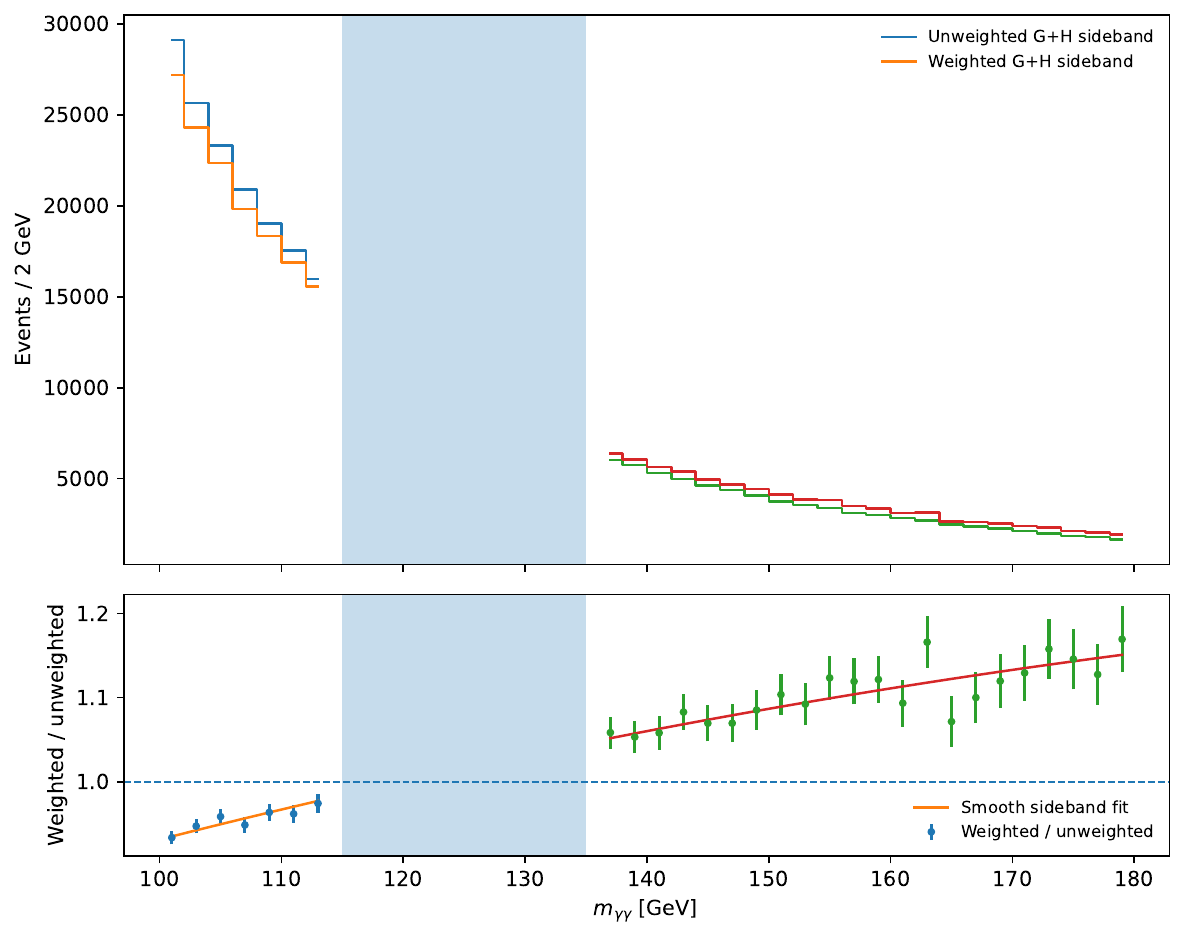}
  \caption{Final application of the frozen weight to the combined Run2016G and Run2016H sidebands. The signal window is fully masked and is not used in this check. Within the observed sidebands, the weighted to unweighted ratio follows a smooth mass dependence without a localized residual.}
  \label{fig:cmssideband}
\end{figure}

After the construction is frozen, the same architecture is rebuilt with all sideband data and the full fixed 300000-event selected signal pool. The weights are normalized to $\langle w\rangle_B=1$ in the sidebands. The final background second moment is 1.996, corresponding to $N_{\rm eff}^B/N_B=0.501$, while the signal mean weight is 2.140. The final sideband application is shown in Fig.~\ref{fig:cmssideband}. Weighting changes the smooth continuum slope; the no sculpting requirement is not that the weighted and unweighted shapes be identical. Within the observed sideband domain, the weighted to unweighted ratio is smooth rather than locally structured. Because the signal window remains masked in this check, this is empirical evidence for smooth behavior, not a direct background-only measurement inside 115--135~GeV. The held out sample supplies the pre-unblinding stability check; the full-sideband rebuild is the final application. A percent-level sideband closure residual is not a bound on signal bias inside the masked window. The practical checks and uncertainty treatments for that extrapolation are described in Sec.~\ref{sec:systematics}.

\subsection{Observed resonance benchmark}
\label{sec:cmsobserved}

The observed fits use 2~GeV mass bins over 100--180~GeV. The sideband events are divided into 16 intervals of the frozen score (equivalently, of its monotonic weight), with boundaries at unweighted sideband score quantiles, giving approximately equal populations (ties can prevent exact equality). In each interval $k$, the unweighted mass counts are fitted with a Poisson likelihood using the bins whose centers lie outside 115--135~GeV, and an exponential of a cubic polynomial. The expected count in mass bin $j$ is
\begin{equation}
 \lambda_{kj}=\exp\!\left[\sum_{\ell=0}^3 a_{k\ell}z_j^\ell\right],\qquad
 z_j=\frac{m_j-140~{\rm GeV}}{40~{\rm GeV}}.
 \label{eq:stratamodel}
\end{equation}
These fits are extrapolated across the masked window. Let $\langle w\rangle_k$ and $\langle w^2\rangle_k$ be the empirical sideband weight moments in stratum $k$. The baseline and weighted background expectations are
\begin{equation}
 b_j^{(0)}=\sum_k\lambda_{kj},\qquad
 b_j^{(1)}=\sum_k\lambda_{kj}\langle w\rangle_k,\qquad
 b_j^{(2)}=\sum_k\lambda_{kj}\langle w^2\rangle_k.
 \label{eq:stratamoments}
\end{equation}
Using stratum-wide weight moments approximates any remaining mass dependence within a narrow score interval. The strata are a technical interpolation device, not the final sensitivity categories being compared.

The baseline selected unweighted spectrum is fitted with a binned Poisson likelihood, with mean $N_S t_{S,j}^{(0)}+\beta b_j^{(0)}$. The weighted spectrum uses Eq.~\eqref{eq:weightedlikelihood} with no calibration constraints in this benchmark and with
\begin{equation}
 \nu_j=N_S t_{S,j}^{(1)}+\beta b_j^{(1)},\qquad
 V_j=N_S t_{S,j}^{(2)}+\beta b_j^{(2)}.
 \label{eq:cmsfitmoments}
\end{equation}
Here $\beta$ is a profiled background normalization. The templates $t_{S,j}^{(q)}$ are computed by histogramming the detector level signal MC with weights $w^q$ and dividing by its total selected event count; in particular $\sum_jt_{S,j}^{(0)}=1$, while $\sum_jt_{S,j}^{(1)}=\kappa_S$. Thus both fits estimate $N_S$ in the same selected event units. The weighted likelihood uses the modeled $V_j$, not the observed $q_j^w$.

For both fits, $N_S\ge0$ in the discovery diagnostic and $Z=\sqrt{2[\ln L(\widehat N_S,\widehat\beta)-\ln L(0,\widehat\beta_0)]}$. The quoted yield errors use the leading mean-template information $I_{ab}=\sum_j(\partial_a\nu_j)(\partial_b\nu_j)/V_j$, with $a,b\in\{N_S,\beta\}$, and $\sigma_{N_S}=(I_{SS}-I_{S\beta}^2/I_{\beta\beta})^{-1/2}$. For the unweighted Poisson fit $V_j=\nu_j$; for the weighted fit this is the high count information approximation, validated with the conditional ensembles below.

\begin{figure}[H]
  \centering
  \includegraphics[width=0.94\textwidth]{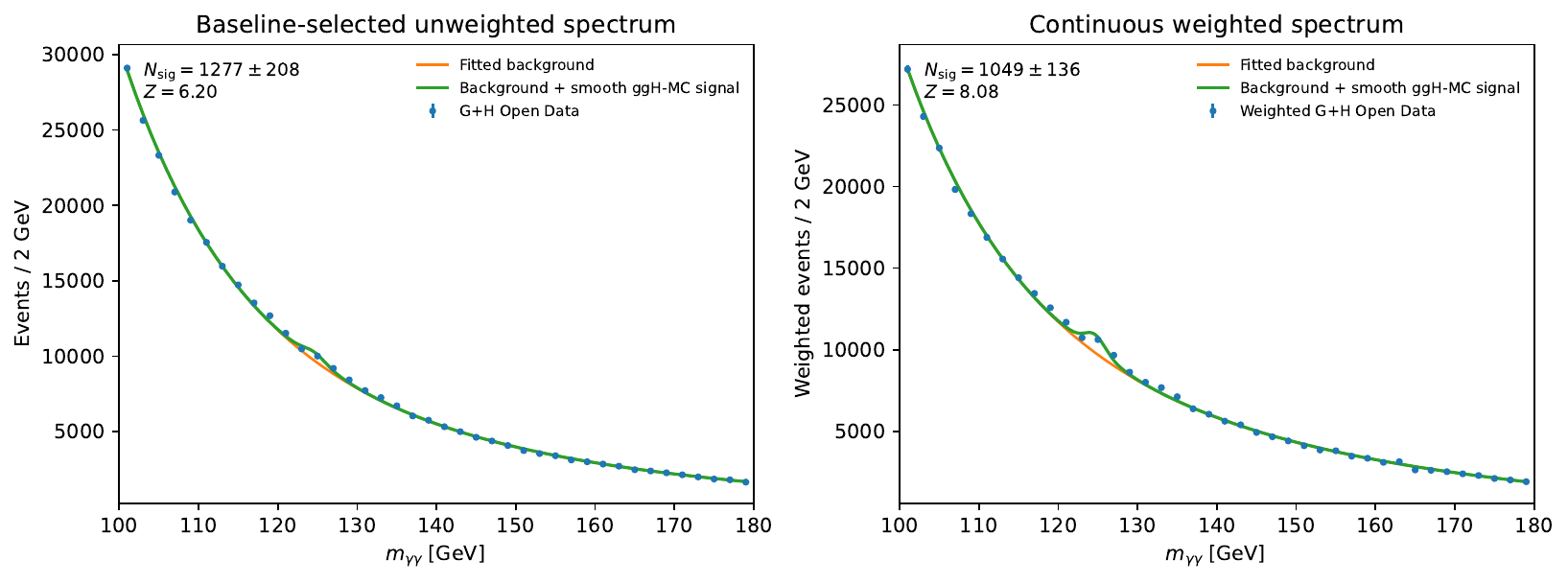}
  \caption{Combined CMS Run2016G and Run2016H Open Data after the common baseline diphoton selection, before and after application of the frozen continuous weight. The smooth background plus signal curve uses a two Gaussian representation fitted to the same detector level gluon fusion MC mass distribution for display only; the numerical likelihood uses the frozen binned detector level MC template. The weight is constructed from sidebands and selected signal MC without using the 115--135~GeV data window. The fitted signal normalization remains in selected event units.}
  \label{fig:cmsdata}
\end{figure}

The numerical signal model is the binned detector level MC mass template and retains its reconstructed tails and asymmetry. The common-mean two Gaussian curve in Fig.~\ref{fig:cmsdata} is only a smooth display representation, not the likelihood template. After the sideband fits, background shapes are fixed apart from $\beta$; finite sideband and background family uncertainties are not propagated into the quoted diagnostic $Z$ values. A simultaneous fit over sidebands and signal region with profiled background shape parameters is a natural alternative for an experimental application, following Sec.~\ref{sec:systematics}. The two-step construction is retained here as a common fixed benchmark for the methods.

After all choices are frozen, the Higgs window is included. Using the detector level MC signal template in the fit to real collision data, the unweighted baseline selected benchmark gives
\begin{equation}
  \widehat N_{\rm sig}^{\rm unw}=1277\pm208,
  \qquad Z_{\rm unw}=6.20,
  \label{eq:rawcms}
\end{equation}
and the continuous weighted benchmark gives
\begin{equation}
  \widehat N_{\rm sig}^{w}=1049\pm136,
  \qquad Z_w=8.08.
  \label{eq:weightedcms}
\end{equation}
Here $N_{\rm sig}$ is the selected signal normalization under the common gluon fusion benchmark template, expressed in event units. The weighted value already uses the calibrated moment template of Eq.~\eqref{eq:weightedtemplates}; it must not be divided by $\kappa_S$ again. The two fitted normalizations are strongly correlated estimates from the same data and are not interpreted as independent rate measurements. Their average yield recovery is checked with the ensemble in Sec.~\ref{sec:cmscomparison}. In particular, no comparison with the inclusive Standard Model Higgs rate is made: this method demonstration does not include the precision acceptance, efficiency, calibration, and systematic uncertainty treatment required for such a measurement. The quoted $Z$ values are local template fit diagnostics for comparing the two representations and are not the official CMS Higgs significance.

The CMS Open Data benchmark demonstrates that auxiliary discrimination can be transferred into a single weighted resonance spectrum without destroying its conventional interpretation. The known 125~GeV structure becomes substantially more prominent, and the fitted selected signal normalization uncertainty decreases from 208 to 136 events. These numbers are local template fit diagnostics rather than a precision CMS Higgs measurement, and yield recovery and uncertainty calibration are tested with the controlled ensembles rather than inferred from the appearance of the observed peak.

\subsection{Hard selection, finite score bins, and continuous weighting}
\label{sec:cmscomparison}

A second ensemble tests different uses of the same frozen auxiliary score. Background events are generated from the sideband only score stratified model of Eqs.~\eqref{eq:stratamodel} and \eqref{eq:stratamoments}. Each stratum has a Poisson fluctuated total with mean $\sum_j\lambda_{kj}$. For each event its mass bin is sampled with probability proportional to $\lambda_{kj}$, its mass is uniform within that 2~GeV bin, and its weight is independently resampled from the empirical sideband weight pool in the same stratum. This approximates the joint mass-score distribution by retaining its between-stratum dependence and neglecting its residual within-stratum dependence. Signal mass and weight pairs are resampled together event by event from the selected gluon fusion simulation. The injected mean signal yield is fixed to 760.98 selected events (approximately 761) solely to set a representative signal to background scale; no rate interpretation is attached to this value. One million independent pseudoexperiments are generated.

The optimized hard selection is chosen from sideband data and signal simulation using the expected profiled mass fit information. To make the finite score comparison nonarbitrary, the 4, 8, and 16 bin schemes are defined by quantiles of the frozen background sideband score, so every bin contains the same expected sideband background fraction. The partitions are therefore nested refinements of the same score ordering and are fixed before pseudoexperiment generation. These are analysis defined bins, not official CMS Higgs categories. In the category fits an independent smooth background normalization is profiled in each score interval, whereas the continuous weighted fit uses one weighted spectrum. Consequently, the finite category pseudoexperiment comparison is a practical category fit benchmark rather than an exact numerical identity with the analytic theorem of Sec.~\ref{sec:categories}.

\begin{figure}[H]
  \centering
  \includegraphics[width=0.80\textwidth]{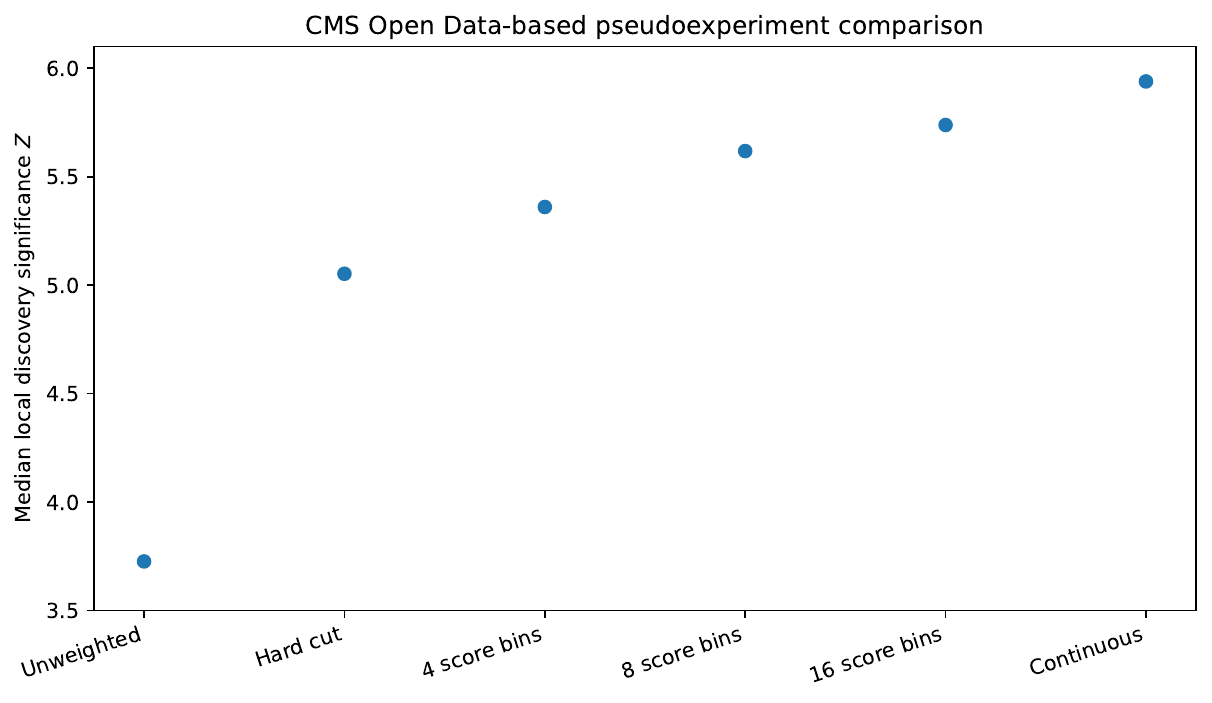}
  \caption{Median local template fit sensitivity in one million CMS Open Data based pseudoexperiments. The 4, 8, and 16 point entries are bins of the same frozen auxiliary score, not official CMS event categories. Error bars denote the Monte Carlo uncertainty on each median and are smaller than the marker size.}
  \label{fig:methods}
\end{figure}

The central numerical demonstration of the discrete to continuous hierarchy can be formed before the mass fit from the expected score fractions. Relative to the unweighted baseline selected spectrum, the simple background dominated gains are 1.315 for the hard cut, 1.411 for 4 score bins, 1.461 for 8 bins, 1.482 for 16 bins, and 1.515 for the continuous weight. The monotonic sequence shows directly how progressively finer use of the same auxiliary information approaches the continuous construction. The full pseudoexperiment fits, which include separate category background normalizations, preserve the same ordering in Table~\ref{tab:methods} and Fig.~\ref{fig:methods}.

\begin{table}[H]
  \centering
  \caption{Expected performance from $10^6$ CMS Open Data based pseudoexperiments. The signal normalization is a benchmark selected event yield. The quoted uncertainty is the median fitted uncertainty on that normalization.}
  \label{tab:methods}
  \begin{tabular}{lcc}
    \toprule
    Method & Median $Z$ & Median $\sigma(N_S)$ \\
    \midrule
    Unweighted selected & 3.727 & 205.66 \\
    Optimized hard cut & 5.053 & 153.51 \\
    4 score bins & 5.360 & 144.65 \\
    8 score bins & 5.618 & 138.53 \\
    16 score bins & 5.738 & 135.96 \\
    Continuous weight & 5.939 & 133.43 \\
    \bottomrule
  \end{tabular}
\end{table}

\begin{figure}[H]
  \centering
  \includegraphics[width=0.66\textwidth]{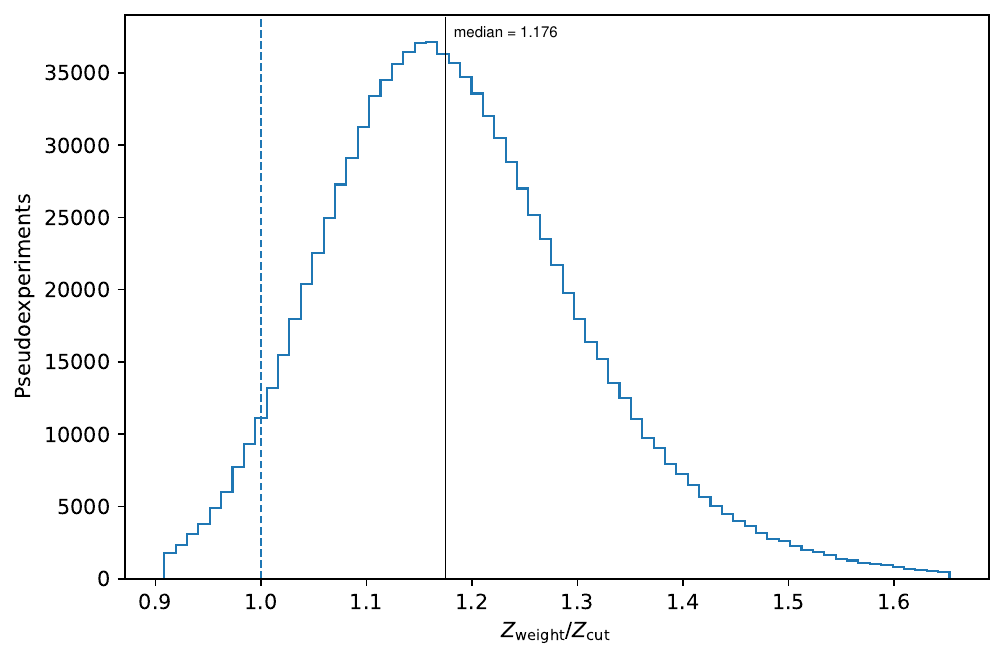}
  \caption{Paired distribution of $R_i=Z_{{\rm weight},i}/Z_{{\rm cut},i}$ in the CMS Open Data based pseudoexperiments. The two methods are evaluated on the same generated experiment in each entry. The dashed vertical line at unity marks equal sensitivity; the second vertical line marks the median, $R_i=1.176$.}
  \label{fig:paired}
\end{figure}

The hard cut and continuous weight are evaluated on the same generated pseudoexperiment, so their comparison can be made pairwise rather than only through the separately quoted medians in Table~\ref{tab:methods}. For each trial $i$ we define
\begin{equation*}
  R_i=\frac{Z_{{\rm weight},i}}{Z_{{\rm cut},i}},
  \qquad
  \Delta Z_i=Z_{{\rm weight},i}-Z_{{\rm cut},i}.
\end{equation*}
The ratio expresses fractional improvement and the difference expresses an absolute change in the diagnostic significance. They satisfy $\Delta Z_i=(R_i-1)Z_{{\rm cut},i}$ for $Z_{{\rm cut},i}>0$ and therefore are related, but have different ensemble distributions when the cut sensitivity fluctuates. The median paired ratio is $\operatorname{median}(R_i)=1.176$, while 95.1\% of the pseudoexperiments satisfy $R_i>1$; the median paired difference is $\operatorname{median}(\Delta Z_i)=0.887$. These paired quantities are distinct from the simple pre-fit sensitivity proxies quoted above. In particular, the value 1.176 is the median of the full-fit paired ratio, not the ratio $1.515/1.315$ of the continuous and hard-cut pre-fit proxies. Figure~\ref{fig:paired} shows the distribution of $R_i$.

Let $L$ denote integrated luminosity. Under the usual statistics-dominated scaling $Z\propto\sqrt{L}$, the paired median sensitivity ratio corresponds to the benchmark luminosity factor
\begin{equation}
  \frac{L_{\rm cut}}{L_{\rm weight}}
  \simeq \left[\operatorname{median}(R_i)\right]^2
  =(1.176)^2\simeq1.38.
  \label{eq:lumiequiv}
\end{equation}
Thus the optimized hard selection would require about 38\% more integrated luminosity to reach the same median sensitivity in this benchmark. This conversion uses the usual square-root luminosity scaling and is specific to the chosen signal and background model.

For each of the unweighted, optimized hard cut, 4/8/16 score-bin, and continuous weight fits, the selected signal normalization has approximately calibrated statistical errors under this fixed model. The ensemble mean estimates are 761.11 events for the unweighted fit and 760.11 for the continuous fit, compared with the common injected mean 760.98. Pull widths across the six methods range from 1.0002 to 1.0010, $1\sigma$ coverage is approximately 0.683 (Gaussian nominal value 0.6827), and the $1.96\sigma$ coverage is approximately 0.950. For the continuous weight the ensemble mean differs from the injected yield by $-0.11$\%, while the pull width is 1.0010. The sensitivity gain is therefore not produced by underestimating the fitted statistical uncertainty.

\section{Implications for resonance searches}
\label{sec:discussion}

The practical use case is a search for a new resonance $X$ with a well defined resonance variable. A signal hypothesis supplies detector level simulated events from which an auxiliary density ratio or calibrated classifier response can be obtained, while background information can come from sidebands, control samples, or simulation. The analysis then follows a compact sequence:
\begin{enumerate}
  \item define the resonance variable $m$ and do not determine or optimize the weight using a localized structure in its signal region;
  \item use auxiliary observables to construct a nonnegative sensitivity weight;
  \item freeze the modeling choices and validate the weighted background against signal-like bias, as described in Sec.~\ref{sec:systematics};
  \item fit a single weighted resonance spectrum with its first and second weight moments included in the statistical model.
\end{enumerate}

This construction is especially useful when several observables provide moderate discrimination but the final experimental object is still intended to be one conventional resonance spectrum. A hard selection discards part of the auxiliary information. A finite set of sensitivity driven bins retains more information but introduces boundaries whose placement is not intrinsic to the resonance physics. The analytic result of Sec.~\ref{sec:categories} identifies those bins as a piecewise constant approximation to the optimal linear weight under its stated assumptions, while the CMS benchmark shows the same information hierarchy numerically on a realistic detector level example.

The method is not intended to erase categories that carry distinct physics content. Categories with different mass resolutions, signal shapes, production parameters, control regions, or nuisance structures can remain physically and statistically useful. Nor does continuous weighting remove the ordinary requirements of detector modeling, correlation treatment, background validation, and systematic uncertainty propagation. The optimality result derived here is the background dominated linear result for a common signal hypothesis; low count problems, very long weight tails, or substantially different likelihood structures may require a more general treatment. The CMS example is correspondingly a real data method demonstration with detector level signal MC and a simplified sideband background model, not a precision Higgs measurement.

Equation~\eqref{eq:joint} assumes an additive nonnegative signal plus background intensity. Resonance signals for which signal background interference is an essential signed contribution require a generalized template treatment and are outside the present formulation.

Within this scope, the methodological statement is direct: sensitivity driven event partitioning is a discretized use of auxiliary information. Continuous weighting removes that discretization while preserving the resonance coordinate for inspection, fitting, and physical yield extraction.

\section{Conclusions}

We have shown that, for a common resonance hypothesis, hard selection and sensitivity driven categorization can be understood as discrete realizations of a continuous event weighting framework. A hard selection is a binary event weight, a finite sensitivity partition is a piecewise constant weight, and in the background dominated linear limit the fine partition approaches the continuous auxiliary signal to background density ratio. This identifies category boundaries as a discretization of information when their primary role is sensitivity rather than distinct physics content.

The construction keeps the resonance spectrum as the experimental observable. Weighted yields and variances are determined by the first and second weight moments, the background model must control localized signal like sculpting, and the fitted coefficient can remain parameterized directly in selected signal event units. One million fully explicit pseudoexperiments with Gaussian, Gamma, and Beta auxiliary observables give pull widths within about one per mille of unity and approximately nominal $1\sigma$ and $1.96\sigma$ coverage while improving the median fitted yield sensitivity from 2.64 to 5.40.

The CMS Run2016G and Run2016H Open Data benchmark provides a complementary test on real collision data. A detector region conditioned weight with $\alpha=0.75$ is frozen from sidebands and detector level gluon fusion Higgs MC before the Higgs window is examined. In the observed benchmark the known 125~GeV structure becomes substantially more prominent: the local template fit diagnostic increases from 6.20 to 8.08 and the fitted selected signal normalization uncertainty decreases from 208 to 136 events. These quantities are used as method diagnostics rather than as a precision Higgs rate or significance measurement.

A separate one million pseudoexperiment study using the same frozen CMS score gives median sensitivities of 5.05 for an optimized hard cut, 5.36, 5.62, and 5.74 for 4, 8, and 16 score bins, and 5.94 for continuous weighting. The ordered progression is the central numerical demonstration that increasingly fine use of the same auxiliary information approaches the continuous construction. Relative to the optimized hard selection, continuous weighting improves the median sensitivity by about 18\%, corresponding in this benchmark to about 38\% more luminosity for the hard selection. The finite score bins are analysis defined intervals of the same frozen score, not the official CMS Higgs categories.

These results establish continuous event weighting as a practical alternative to sensitivity driven event partitioning for resonance searches in which a common signal hypothesis is extracted from a single resonance spectrum. The method retains the information recovered by increasingly fine categorization without requiring increasingly elaborate sensitivity boundaries, while preserving direct access to the physical selected signal yield and its statistical uncertainty. For future searches, this provides a natural route from one or a small number of detector level signal simulations and data driven background control samples to an information efficient but experimentally transparent resonance analysis.

\end{document}